\documentclass[12pt]{iopart}

\usepackage{bm}                     
\usepackage{appendix}
\usepackage[latin1]{inputenc}
\usepackage{dcolumn}      
\usepackage{bbm}
\usepackage{color}
\usepackage{xcolor}
\usepackage{esint}

\usepackage[mathscr]{eucal}
\usepackage{latexsym}
\usepackage{amsbsy}
\usepackage{float}
\usepackage{graphicx}
\usepackage{epsfig}
\usepackage{subfigure}
\usepackage{wrapfig}
\usepackage{epsf}
\usepackage{float}
\usepackage[normalem]{ulem}
\usepackage{qcircuit}

\usepackage{amsfonts}

\definecolor{darkblue}{HTML}{004D6B}
\definecolor{darkred}{HTML}{8c1515}
\definecolor{darkgreen}{HTML}{006400}
\usepackage{parskip}

\newcommand{\be}{\begin{equation}}
\newcommand{\ee}{\end{equation}}
\newcommand{\ba}{\begin{array}{l}}
\newcommand{\ea}{\end{array}}
\newcommand{\re}[1]{(\ref{#1})}
\newcommand{\ci}[1]{\cite{#1}}
\newcommand{\banonum}{\begin{eqnarray*}}
\newcommand{\eanonum}{\end{eqnarray*}}
\newcommand{\baa}{\begin{eqnarray}}
\newcommand{\eaa}{\end{eqnarray}}
\newcommand{\bfr}{\begin{flushright}}
\newcommand{\efr}{\end{flushright}}
\newcommand{\bfl}{\begin{flushleft}}
\newcommand{\efl}{\end{flushleft}}
\newcommand{\lab}[1]{\label{#1}}

\begin{document}

\title[Dirac particle under dynamical confinement]{Dirac particle under dynamical confinement:\\ Fermi acceleration, trembling motion and quantum force}
\author{J. Dittrich$^a$, S. Rakhmanov$^b$ and D. Matrasulov$^c$}
\address{$^a$Nuclear Physics Institute, Czech Academy of Sciences, 25068 \v{R}e\v{z}, Czech Republic \\ $^b$Chirchik State Pedagogical University, 104 Amur Temur Str., 111700 Chirchik, Uzbekistan\\ $^c$Turin Polytechnic University in Tashkent, 17 Niyazov Str., 100095 Tashkent, Uzbekistan}

\begin{abstract} 
Quantum dynamics of a Dirac particle in a 1D box with moving wall is studied. Dirac equation with time-dependent boundary condition is mapped onto that with static one, but with time-dependent mass.
Exact analytical solution of such modified Dirac equation  is obtained for massless particle. For massive particle the problem is solved numerically.
Time-dependences of the main characteristics of the dynamical confinement, such as average kinetic energy and quantum force are analyzed. It is found that  the average kinetic energy remains bounded for 
the interval length bounded from below, in particular for the periodically oscillating wall.
\end{abstract}

\section{Introduction}

Confined quantum systems appear in many topics, such as quantum  materials, quantum optics, quantum information and quantum technologies. Remarkable feature of such systems is the dependence of its physical properties on the size and geometry of confinement. Such a feature can be used as an effective tool for tuning of the functional properties of a quantum material and device optimization. Underlying reason causing this feature is the fact that the geometry and size of the confinement appear in the boundary conditions to be imposed for quantum mechanical wave (e.g., Schr\"{o}dinger, Dirac) equations. Thus confinement geometry which determines the boundary conditions, can manifest on the physical properties of a confined quantum system. Such a feature is especially important for the case of dynamical confinement, i.e., when the boundary conditions determined by the confinement are time-dependent. In this case, besides geometry and size of the confinement, one can use additional tool, i.e. time-dependence (movement law) of the confinement boundary for controlling the quantum dynamics in the confinement area.
Simplest case of the dynamical confinement is an one-dimensional box with one moving wall. We note that the dynamical confinement  described in terms of the Schr\"{o}dinger equation with time-dependent boundary conditions (mostly in 1D case) was subject extensive research earlier  (see, e.g., Refs.\ci{doescher}-\ci{Uzy22}).

Early treatments of the dynamical confinement in nonrelativistic qantum mechanics date back to Doescher, who studied basic aspects of the problem \cite{doescher}. Munier et al.\ studied  more detailed research of the problem and caclculated physically observable quantities for the problem of time-dependent box \cite{mun81}. Later, Makowsky \cite{mak91}-\cite{mak923} and Razavy \cite{razavy01,razavy02} provided  a systematic study of the dynamical confinement problem, by considering one-dimensional box with moving walls and classifying time-dependence of the wall approving exact solution of the Schr\"{o}dinger equation with time-varying boundary conditions.  Unitary transformation that maps the time-dependent box to that with fixed walls, was found in \cite{razavy01,razavy02}. These studies used an approach developed earlier by Berry and Klein \cite{berry00}. Different aspects of the problem dynamical confinement and its applications to dynamical Casimir effect was studied in a series of papers by Dodonov et al.\ \cite{dodonov01}-\cite{dodonov04}. Berry phase in time-dependent box was studied in \cite{per,Kwon,Wang}. \v Seba studied the problem of time-dependent box in the context of quantum Fermi acceleration \cite{seb90}. The quantum gas under the dynamical confinement was considered in \cite{nakamura01,nakamura02}, where quantum force operator for time-dependent box was introduced. 

Hydrogen-like atom confined in a spherical hard-wall box with time-varying radius was considered in \cite{Our03}. Symmetry properties of the time-dependent harmonic oscillator reproduced by a time-dependent quantum box was presented in a series of papers by Lewis \cite{lewis01,lewis02}. Different aspects of the problem of dynamical confinement were studied also in \cite{pinder} -\cite{glas09}. Inverse problem for time-dependent box, i.e.\ the problem of recovering boundary's time-dependence from existing solution is considered in \ci{james}. Dynamical confinement in a half-line has been considered in \ci{smith}. The problem of time-dependent Neumann boundary conditions is considered in \ci{duca}. Extension of the dynamical confinement to relativistic case   by considering Dirac equation for time-dependent box  was done in \cite{sobirov}. Time-dependent quantum graphs have been considered in the  Refs.\cite{matrasulov,nikiforov,Uzy22}.  Despite the considerable progress made in the study of dynamical confinement for different low-dimensional systems, all or most of the studies are restricted by nonrelativistic systems, although one can mention a paper dealing with classical Klein-Gordon equation with the dynamical boundary conditions \cite{Ditt3}. Extension of such studies for Dirac fermions is still open problem. Motivation for the study of dynamical confinement in Dirac system comes from such practically important problems as MIT bag model \ci{MITBag}, graphene quantum dots \ci{GrapheneDot}, optical Dirac equations \ci{Zeuner}-\ci{Silva} and other problems of the relativistic optics, where confinement boundary always undergo some fluctuations. This paper is organized as follows. In the next section we will give brief description of the problem of dynamical confinement for the Schr\"{o}dinger equation. Section III presents formulation and solution of the problem of dynamical confinement for Dirac particle in a time-dependent box.  In section IV we present analysis of expectation values of kinetic energy, quantum force, and particle's position as a function of time. Finally the section V provides some concluding remarks.

\section{Nonrelativistic counterpart:\\ Schr\"{o}dinger equation for 1D time-dependent box}

Dynamics of a quantum particle in time-dependent box, i.e. in a hard-wall box with moving (one) wall is described in terms of the following  time-dependent Schr\"{o}dinger equation ($\hbar = m = 1$):
\be 
i\frac{\partial \Psi(x,t)}{\partial t}=-\frac{1}{2}\frac{\partial^2}{\partial x^2}\Psi(x,t),
\quad t\in [0,\infty), x\in [0, L(t)]
\lab{shr0001}
\ee
For such system, the boundary conditions are imposed as \ci{mak91}
\begin{equation*}
\Psi (x,t)|_{x=0}=\Psi (x,t)|_{x=L(t)}=0. \nonumber
 \label{bc} 
\end{equation*}

Using the transformations for coordinate and for solution given as
\baa 
y=\frac{x}{L(t)}, {\rm and}\;\; \Psi(x,t)=\sqrt{1/L}\exp{\Biggl( \frac{i}{2}L\dot{L} y^2
\Biggl)}\varphi(y,t), \lab{trans002} \nonumber
\eaa
Eq. \re{shr0001} can be rewritten as \ci{mak91}
 \be
 i\frac{\partial\varphi(y,t)}{\partial
t}=-\frac{1}{2L^2}\frac{\partial^2 \varphi(y,t)}{\partial y^2}+
 \frac{1}{2} L\ddot{L}y^2 \varphi(y,t),
\lab{shr002}
\ee
where $\ddot{L}=d\dot{L}/dt$ and $\varphi(y,t)$ satisfies the boundary conditions given by
\baa 
\varphi(y,t)|_{y=0}=\varphi(y,t)|_{y=1}=0. \nonumber
\eaa
Space- and time-variables in Eq.\re{shr002} can be separated provided $L(t)$ fulfills the condition $L^3\ddot{L} =\textnormal{const}.$ Different exactly solvable cases following from this conditions have been considered in \ci{mak91}. For arbitrary $L(t)$ which does not fulfill the condition for factorisation, Eq.\re{shr002} should be solved numerically.

Alternative approach for the problem of time-dependent boundary conditions in the one-dimensional Schr\"{o}dinger equation was proposed in \ci{Pere_Pro} via introducing extended version of the time-derivative in the time-dependent Schr\"{o}dinger equation Namely, by introducing an operator $\hat P$, describing the evolution of the wave function corresponding to the evolution of (time-dependent) interval, [0, L(t)],  such that $[0, L(t_1)] \to [0, L(t_2)]$
and
\begin{eqnarray*}
\hat P: L^2([0,L(t_1)]) \to  L^2([0,L(t_2)]): \Psi(\cdot,t_1) \mapsto \Psi(\cdot,t_2),
\\
\Psi(x,t_1) = \sqrt{\frac{L(t_1)}{L(t_2)}}\Psi\left(x \sqrt{\frac{L(t_2)}{L(t_1)}},t_2\right) \quad,\quad
x \in [0,L(t_1)]
\end{eqnarray*}
Since the time-dependent interval, $[0, L(t)]$ is parametrized by coordinate $x$, its action on this interval can be realized as dilatation
in coordinate representation, i.e., 
\baa
x \mapsto x_1 = \frac{x L(t_1)}{L(t)}. \nonumber
\eaa
Therefore, the action of $\hat P$ on $\Psi(x,t)$ in the infinitesimal form can be written as \ci{Pere_Pro}
\baa 
\hat{P}=1+\delta t \frac{\dot{L}}{2L}(x\partial_x+\partial_xx) \nonumber
\eaa
Then the time derivative of the wave function for time-varying boundaries can be determined as \ci{Pere_Pro}
\baa 
\nabla_t\Psi(x,t)= \lim_{\delta t\to 0}  \frac{1}{\delta t}[ \hat{P}(t,t+\delta t)\Psi(x,t+\delta t)-\Psi(x,t)] = \nonumber
\eaa
\be
\left[\partial_t+\frac{\dot{L}}{2L}(x\partial_x+\partial_xx)\right]\Psi(x,t)
\lab{modified}
\ee
It is clear that $\nabla_t$ coincides with $\partial_t$ in the limit of
time-independent boundary, i.e., at $\dot{L}=0$. It takes into account also the drift caused by the 
spatial support $[0,L(t)]$ motion.

Thus in terms of the modified time-derivative, the Schr\"{o}dinger equation describing motion of a quantum particle in 1D box with moving wall can be written as
\baa 
i\nabla_t\Psi(x,t)=-\frac{1}{2}\partial_{xx}\Psi(x,t) \nonumber
\eaa 
or
\be 
i\partial_t\Psi(x,t)=-\frac{1}{2}\partial_{xx}\Psi(x,t)-i\frac{\dot{L}}{L}\left(x\partial_x+\frac{1}{2}\right)
\lab{schr1}
\ee
Using transformations of the coordinate, time and wave function given by
\baa
y=\frac{x}{L}, \;\;\; \tau=\int_0^t\frac{ds}{[L(s)]^2}, \;\;\; \Psi(x,t)=\frac{1}{\sqrt{L}}\Psi_1(y,\tau) \nonumber
\eaa
Eq.\re{schr1} can be written as
\baa 
i\partial_\tau\Psi_1=-\frac{1}{2}\partial_{yy}\Psi_1 \nonumber
\eaa
Eigenfunction solutions of this equation are given as
\baa 
\Psi_{1,n}(y,\tau)=\sqrt{2}e^{-i\frac{\pi^2n^2}{2}\tau}\sin(\pi n y), \nonumber
\eaa
Thus, for the solution of Eq. \re{schr1} we have
\baa 
\Psi_n(x,t)=\sqrt{\frac{2}{L}}e^{-i\frac{\pi^2n^2}{2}\tau(t)}\sin\frac{\pi n x}{L} \nonumber
\eaa
or
\baa 
\Psi_n(x,t)=\sqrt{\frac{2}{L}}e^{-i\frac{\pi^2n^2}{2}\int_0^t\frac{ds}{[L(s)]^2}}\sin\frac{\pi n x}{L}. \nonumber
\eaa
We note that this solution is valid for arbitrary $L(t)$.

\section{Dirac particle under dynamical confinement}
The problem of the quantum time-dependent confinement can be studied  also  for relativistic systems, by considering, e.g., one-dimensional time-dependent Dirac equation in a hard wall box with moving wall. Earlier, the problem of dynamical confinement for Dirac particle was considered in \ci{sobirov} by considering the problem of time-dependent neutrino billiard. The main problem arising in the study of Dirac equation with time-dependent boundary conditions is the breaking of the norm conservation. In \ci{sobirov}  such problem was avoided in special case of massless Dirac  particle confined in a box with linearly moving wall. Treatment of quantum dynamics of a spin-half particle in a hard wall box is rather complicated problem due to the possible particle-antiparticle pair creation in strong fields. Therefore one needs to impose the problem in terms of the boundary conditions by assuming that these latter are not caused by strong external field. Such a situation appears, e.g., in graphene quantum dots \ci{Peeters} and different versions of Dirac billiards \ci{DB2,DB3}. Strict formulation of the  problem of a Dirac particle in a hard wall box was presented in a pioneering paper by Berry and Mondragon \ci{Berry02}. The case of relativistic particle in one-dimensional box was considered in a series of papers by Alonso, et.al. \ci{Alonso,Alonso02}. Different versions of a system Dirac particle + 1D box have been considered on the context of graphene quantum dots \ci{GrapheneDot}. Before starting the discussion of the problem of Dirac particle under dynamical confinement, following to the Ref. \ci{Alonso}, we briefly recall the problem of Dirac particle under static confinement caused by one-dimensional hard wall box.

\subsection{Dirac particle in 1D box: Static confinement}

In the absence of external potential, time-independent Dirac equation can be written as ($\hbar=c=1$)

\begin{equation}
    H_0\psi=(-i\alpha\nabla+m\beta)\psi=\varepsilon\psi \label{eq01}
\end{equation}

where $\alpha$, $\beta$ are the Dirac matrices,

\begin{equation*}
    \psi=\left(\begin{array}{c}\phi\\\chi\end{array}\right) \nonumber
\end{equation*}

Eq. (\ref{eq01}) is equivalent to the following coupled
equations:
\begin{equation*}
    -i\cdot\frac{d\chi}{dx}+m\phi=\varepsilon\phi \nonumber
\end{equation*}

\begin{equation*}
    -i\cdot\frac{d\phi}{dx}-m\chi=\varepsilon\chi. \nonumber
\end{equation*}

Solution of Eq. (\ref{eq01}) for the boundary condition $\phi(0)=\phi(L)=0$ can be written as
\begin{equation*}
    \psi=A\left(\begin{array}{c}i\sin(kx)\\ \frac{k}{\varepsilon+m}\cos(kx)\end{array}\right) \label{wf01} \nonumber
\end{equation*}
with $k=n\pi/L, n=1,2,...$ and one can find energy levels with $\varepsilon=\sqrt{k^2+m^2}$.
The boundary condition, $\chi(0)=\chi(L)=0$ yields the eigenfunctions
\begin{equation*}
    \psi=A\left(\begin{array}{c}\cos(kx)\\ \frac{ik}{\varepsilon+m}\sin(kx)\end{array}\right) \label{wf02}. \nonumber
\end{equation*}
In the next section, these eigenfunctions will be used to construct solution of the time-dependent Dirac equation in a time-varying interval.

\subsection{Dirac particle in a time-dependent box}

Consider the time-dependent Dirac equation in 1+1 dimensional space-time 
\begin{equation}
\label{DSchr}
i\partial_t \Psi = H\Psi := (-i \alpha \partial_x +\beta m)\Psi .
\end{equation}

We assume that particle is trapped inside the box with moving wall and the position of the wall is given by function $L(t)$. In such formulation the problem becomes relativistic counterpart of the model considered in section II. For the Dirac spinor $\psi$ and the Dirac matrices given by
\begin{equation*}
\label{alpha_beta}
\Psi=\left(\begin{array}{c}\psi_1\\\psi_2\end{array}\right), \;\;\;
\alpha=
\left(
\begin{array}{rr}
0 & 1\\
1 & 0
\end{array}
\right)
\quad,\quad
\beta=
\left(
\begin{array}{rr}
1 & 0\\
0 & -1
\end{array}
\right)
, \nonumber
\end{equation*}
 in the interval $[0,L(t)]$, the motion of the particle is described by Eq.\re{DSchr} for which the boundary conditions are imposed as
\begin{equation*}
\label{bound_cond}
{\mathcal D}(H)=\{ \Psi \in H^1((0,L(t)), {\mathbb C}^2) | \psi_1(0)=\psi_1(L(t))=0 \}. \nonumber
\end{equation*}
It is easy to check that these boundary conditions  guarantee self-adjointness of the Dirac Hamiltonian $H$.
From the one-parametric family of boundary conditions at each end-point, we choose just one possibility
corresponding to the first boundary condition mentioned in Subsection A.

We should note that for the above boundary conditions the norm conservation for the system described by Eq.\re{DSchr} is broken which means that the probability of finding the particle in $(0,L(t))$ is not conserved:
\begin{equation*}
\partial_t \int_0^{L(t)} \Psi(x,t)^\dag \Psi(x,t) \, dx \;=\; \dot{L}(t) |\Psi(L(t),t)|^2 \neq 0. \nonumber
\end{equation*}

To avoid such unphysical situation, in  Eq. \re{DSchr}, we replace 
the time-derivative with the modified one determined by Eq. \re{modified}, i.e use at the left hand side of  Eq. \re{DSchr} time-derivative
\begin{equation*}
\nabla_t=\partial_t + \frac{\dot{L}(t)}{2 L(t)}(x \partial_x +\partial_x x).
\lab{modnabla} \nonumber
\end{equation*}

The modified Dirac equation now reads
\begin{equation}
\label{DirMod}
i\nabla_t \Psi = (-i \alpha \partial_x +\beta m)\Psi .
\end{equation}
It is easy to check that Eq. \re{DirMod} provides conservation of probability:
\begin{equation*}
\frac{d}{dt}\int_0^{L(t)} \Psi(x,t)^\dag \Psi(x,t) \, dx =0. \nonumber
\end{equation*}

Furthermore, we use the following transformations of coordinate, time and the wave function:
\begin{equation}
y=\frac{x}{L(t)}. \;\;\; \tau=\int_0^t \frac{1}{L(\xi)}d\xi, \;\;\; 
\psi(x,t) = \frac{1}{\sqrt{L(t)}} \varphi(y,\tau(t)),\; 
\lab{trans1}
\end{equation}
with $ \varphi=\left(\begin{array}{c}\varphi_1\\\varphi_2\end{array}\right).$
Then from Eq. \re{DirMod} we get
\begin{equation}
i\partial_\tau \varphi(y,\tau)=(-i\alpha\partial_y + m L(t(\tau)) \beta ) \varphi(y,\tau)
\lab{DirMod1}
\end{equation}
where $y\in (0,1)$ and $\varphi_1(0,\tau)=\varphi_1(1,\tau)=0$.
Thus we "mapped" the problem onto the Dirac equation on a fixed interval but with a time-dependent mass.

The solutions of (\ref{DirMod1}) can be written in terms of the complete set of the eigenfunctions of the massless Dirac equation by given by 
\begin{eqnarray*}
-i\alpha\partial_y \psi_n = n\pi \psi_n \quad,\quad \psi_{n,1}(0)=\psi_{n,1}(1)=0 , \nonumber
\end{eqnarray*}
where
\be
\psi_n(y) = \left(
\begin{array}{c}
\sin(n\pi y) \\- i \cos(n\pi y)
\end{array}
\right) , \;\;\; (\psi_n,\psi_k )=\delta_{n,k}.
\lab{eigen1}
\ee

Inserting the expansion
\begin{equation}
\label{phi_expansion}
\varphi(y,\tau)=\sum_{n=-\infty}^{+\infty} a_n(\tau) \psi_n (y)
\end{equation}
into (\ref{DirMod1}) and projecting onto $\psi^{(n)}$ we obtain
\begin{equation*}
i \dot{a}_n(\tau) = n\pi a_n(\tau) -m L(t(\tau)) a_{-n}(\tau) \quad,\quad n\in {\mathbb Z}, \nonumber
\end{equation*}
that leads to differential equation for $a_0$
\begin{equation}
\label{a_0_eqn}
i \dot{a}_0 = - m L(t(\tau)) a_0
\end{equation}
and the system of ordinary differential equations for $a_n, a_{-n}$
\begin{eqnarray*}
\nonumber
i \dot{a}_n(\tau) = n\pi a_n(\tau) -m L(t(\tau)) a_{-n}(\tau),
\\
 i \dot{a}_{-n}(\tau) = -n\pi a_{-n}(\tau) -m L(t(\tau)) a_{n}(\tau) ,
 \lab{ode2}
\\
\nonumber
n \in {\mathbb N} .
\end{eqnarray*}

Solution of Eq. (\ref{a_0_eqn}) can be written as
\begin{equation}
a_0(\tau)=a_0(0) \exp\left(i m \int_0^\tau L(t(\xi)) \,d\xi \right) = a_0(0) e^{i m t} .
\end{equation}

The system of differential equations \re{ode2} can be written in matrix form as
\begin{eqnarray*}
\label{a_eqn}
i \dot{a}^{(n)}(\tau) =
\left(
\begin{array}{cc}
n\pi & -m L(t(\tau))
\\
-m L(t(\tau)) & -n\pi
\end{array}
\right) a^{(n)}(\tau) ,
\\
a^{(n)}(\tau)=
\left(
\begin{array}{c}
a_n(\tau) \\ a_{-n}(\tau)
\end{array}
\right)
\quad,\quad
n \in {\mathbb N}. \nonumber
\end{eqnarray*}

The equation (\ref{a_eqn}) clearly has a solution which is unique to the initial value $a^{(n)}(0)$ and satisfies
\begin{equation}
|a^{(n)}(\tau)|^2 = |a^{(n)}(0)|^2 . 
\label{a_norm}
\end{equation}
In particular, coefficients $a_{\pm n}$ remain bounded in $\tau$ for every $n \in {\mathbb N}$. For massles Dirac particle, i.e. for $m = 0$,  the system \re{ode2} splits into two uncoupled equations, whose solutions are given as
\be
a_{\pm n}(\tau) =  a_{\pm n}(0)e^{\mp i \pi n \tau}. 
\lab{massless1}
\ee
For linearly moving wall, determined by $L(t) = A +Bt$, by taking into account relation between $\tau$ and $t$ given in Eq.\re{trans1}, one can write explicit solution for massless case as
\baa
a_{\pm n}(t) =  a_{\pm n}(0)e^{\mp i \frac{\pi n}{B} \ln\left(1+\frac{B}{A}t\right)}. \nonumber
\eaa
Solution of the original Dirac equation for time-dependent box given by \re{DirMod} can be found using the relation between $\psi$ and $\varphi$ in Eq.\re{trans1} that yields:
\baa
\Psi(x,t) = \frac{1}{\sqrt{L(t)}} \sum_{n=-\infty}^{+\infty} a_{ n}(0)e^{-i \frac{\pi n}{B} \ln\left(1+\frac{B}{A}t\right)} \psi_n\left(\frac{x}{L(t)}\right), \nonumber
\eaa
where $\psi^{n}$ are given by Eq.\re{eigen1} and $a_{\pm n}$ fulfill the relation following from the norm conservation:
\baa
\sum_{n=-\infty}^{+\infty} |a_{n}(0)|^2 =1. \nonumber
\eaa
Initial values of the expansion coefficients, $a_n(0)$ can be computed from the initial conditions. If one chooses, e.g., the initial condition as a Gaussian wave packet given by \ci{DM18}
\baa
\Psi(x,0) = \frac{f(x)}{\sqrt{s_1^2 +s_2^2}} \left(\begin{array}{c}s_1\\ s_2\end{array}\right) \nonumber
\eaa
with $s_1$ and $s_2$ being the initial spin polarization and
\baa
f(x) = \frac{1}{\sqrt{d\sqrt{2\pi}}}\exp\left[-\frac{(x-x_0)^2}{4d^2} + iv_0 x \right]. \nonumber
\eaa
Here $d,\;x_0$ and $v_0$ are the packet's width, initial position of the center of mass and initial velocity, respectively.
For such initial condition, $a_n(0)$ can be computed as
\baa
a_n(0) =\frac{1}{\sqrt{L(0)}}\int_{0}^{L(0)} \psi_n^{\dag}(x/L(0)) \Psi(x,0) dx.
\lab{initial1} \nonumber
\eaa
where $\psi_n$ are given by Eq. \re{eigen1}.

\section{Fermi acceleration and trembling motion}

\begin{figure}[t!]
\centering
\includegraphics[totalheight=0.3\textheight]{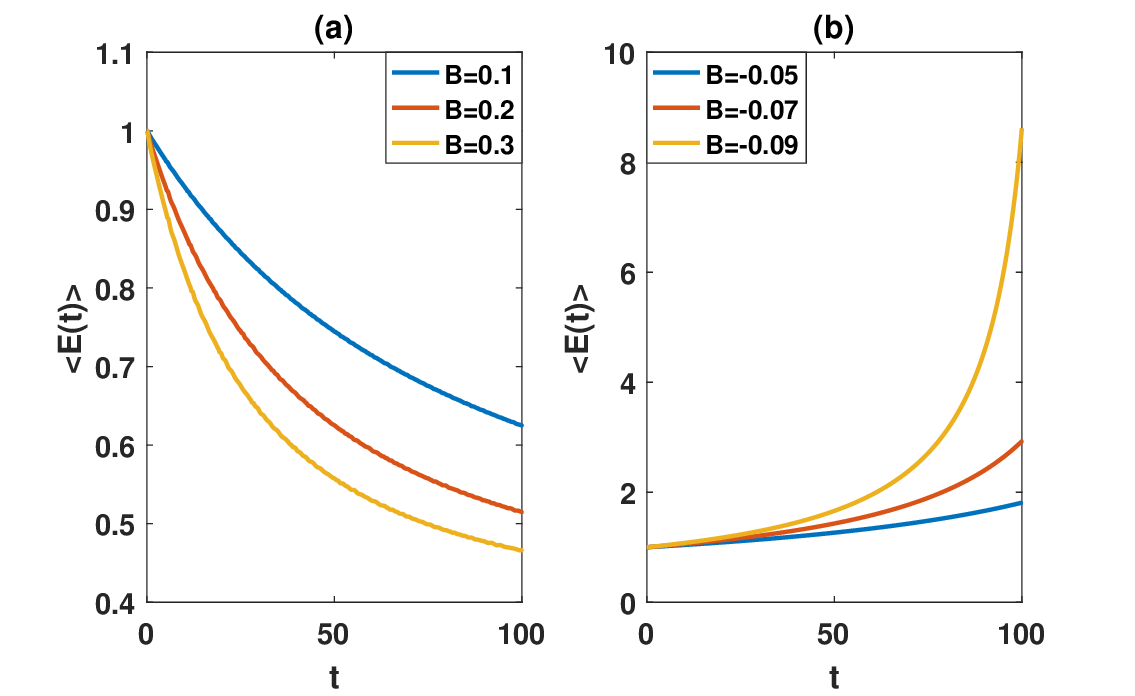}
 \caption{The average kinetic energy of a Dirac particle confined in a  box with linearly moving wall ($L(t)=A+Bt$) as a function of time at different values of the  wall's velocity, $B$ for $m=1$. Two regimes are considered: Expanding (a) and contracting box (b) for $A=10$. The wave packet's width, initial position of the center of mass, initial velocity and the initial spin polarization are chosen as $d=0.1$, $x_0=L(0)/2$, $v_0=0$, $s_1=1$ and $s_2=0$, respectively. }
 \label{fig1}
\end{figure}

One of the features of the time-dependent box is so-called Fermi acceleration, which implies growth of the particle velocity caused by its interaction with the periodically oscillating wall of the box.
An important physically observable characteristics of  the Fermi acceleration in quantum  regime is the average kinetic energy of the particle which is determined as
\baa
<E(t)> = \int_0^{L(t)} \Psi(x,t)^\dag H\Psi(x,t)\, dx . \nonumber
\eaa
Using Eqs.\re{trans1} and \re{phi_expansion} for the average kinetic energy we have
\baa
<E(t)> = \sum_{n=0}^\infty E_n(t) \quad, \nonumber
\eaa
\be
E_0(t) = -m |a_0(\tau(t))|^2 \quad,
\ee
\baa
E_n(t)=\frac{n\pi}{L(t)} (|a_{n}(\tau(t))|^2 - |a_{-n}(\tau(t))|^2 ) - 
\nonumber
\eaa
\be 
2 m \Re (\overline{a_n(\tau(t))} a_{-n}(\tau(t))) 
\quad {\rm for} \; n\in {\mathbb N} .
\lab{energy1}
\ee
Taking into account (\ref{a_norm}), $<E(t)>$ remains bounded in $t$ for $\Psi(0)$ in the domain of the Hamiltonian and $L(t)$ uniformly bounded from below. In other words, no unlimited Fermi acceleration occurs in this case.

For the massless particle, Eqs.\re{energy1} and \re{massless1} lead to
\baa
E_n(t) =
\frac{n\pi}{L(t)} (|a_{n}(\tau(t))|^2 - |a_{-n}(\tau(t))|^2 )=\frac{n\pi}{L(t)} (|a_{n}(0)|^2 - |a_{-n}(0)|^2 ).
\lab{energy2} \nonumber
\eaa
for $n\in {\mathbb N}$.

\begin{figure}[t!]
\centering
\includegraphics[totalheight=0.3\textheight]{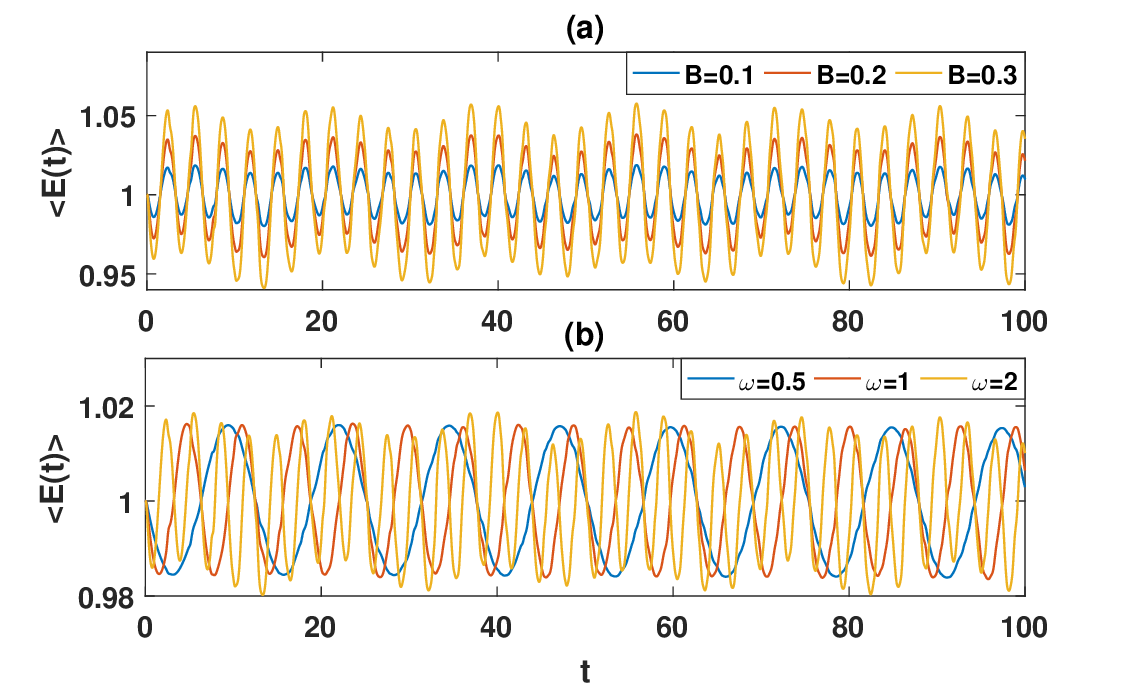}
 \caption{Time-dependence of the average kinetic energy of a Dirac particle in a box with oscillating wall ($L(t)=A+B\sin\omega t$) at different values of amplitude (a)  and frequency (b) of the wall's oscillation for fixed $A=5$ for $m=1$. Parameters are chosen as $\omega=2$ (a) and $B=0.1$ (b). The wave packet's width, initial position of the center of mass, initial velocity and the initial spin polarization are chosen as $d=0.1$, $x_0=L(0)/2$, $v_0=0$, $s_1=1$ and $s_2=0$, respectively.}
 \label{fig2}
\end{figure}

For massive case ($m\neq 0$), the average energy needs to be computed numerically. In Fig.1   plots of the average kinetic energy (as a function of time) are presented for linearly  expanding (a) and contracting  (b) boxes at different values of the wall's velocity. Fig. 2 compares time-dependence of the average kinetic energy of a Dirac particle confined in a box with harmonically oscillating wall at different values of the oscillation amplitude (a) and oscillation frequency (b). In both cases $<E(t)>$ is periodic in time and as higher oscillation amplitudes of the wall, is higher the amplitude of $<E(t)>$.
Increasing of the wall's oscillation frequency causes 
decreasing
of the period of $<E(t)>.$ Thus, no monotonic growth of average kinetic energy is possible for Dirac particle in harmonically oscillating box.

One of the unusual effects in the  Dirac particle's dynamics is so-called "Zitterbewegung", i.e., manifestation of the trembling motion of the particle. It was studied first by Schr\"{o}dinger in \ci{Schrod}, whose results were reproduced later in \ci{Barut}. The effect attracted much attention in different contexts (see, e.g., Refs.\ci{David}-\ci{Sedov}). It was found by considering wave packet motion described by time-dependent Dirac equation, that the average coordinate oscillates as a function of time \ci{Schrod,Barut}. Here we consider this phenomenon for Dirac particle under dynamical confinement provided by time-dependent box.

The main characteristics of Zitterbewegung, average coordinate is determined as 
\baa
<x(t)> = <\Psi(x,t)|x|\Psi(x,t)>. \nonumber
\eaa

\begin{figure}[t!]
\centering
\includegraphics[totalheight=0.3\textheight]{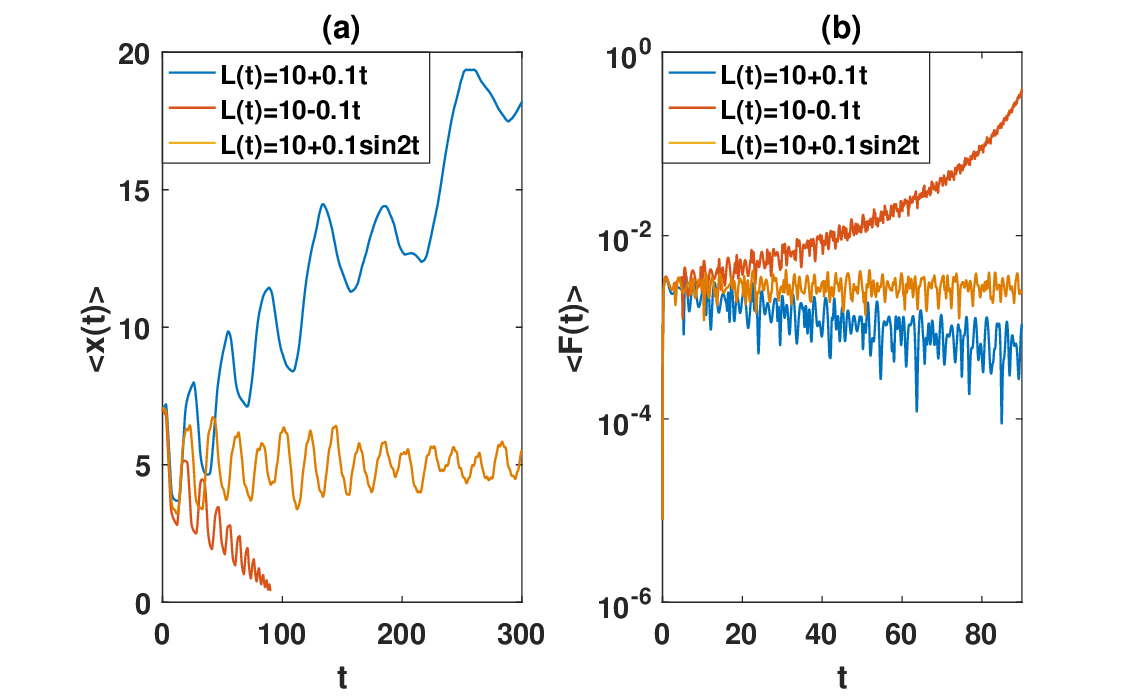}
 \caption{Time-dependent of the average coordinate (a) and force (b) of a Dirac particle in a box with three regime of wall's motion ($L(t)=10+0.1t$, $L(t)=10-0.1t$ and $L(t)=10+0.1\sin 2t$ for $m=1$. The wave packet's width, initial position of the center of mass, initial velocity and the initial spin polarization are chosen as $d=0.1$, $x_0=7$, $v_0=0$, $s_1=1$ and $s_2=0$, respectively.}
 \label{fig3}
\end{figure}
Using Eq.\re{phi_expansion} $<x(t)>$ can be written as
\baa
 <x(t)> =\sum_{n,m} a_n^*(t)a_m(t) V_{nm}, \nonumber
\eaa
where the matrix element $V_{nm}$ can be explicitly written as  

\baa
V_{nm}=\int_0^1y\cos\left(\pi (n-m) y\right)dy=\left\{
\begin{array}{c}
\frac{1}{2}, \;\;\;\;\;\; n=m \\
 \frac{(-1)^{n-m}-1}{\pi^2(n-m)^2}, \;\;\; n\ne m
\end{array} \nonumber
\right.
\eaa
For massles Dirac particle, i.e. for $m = 0$, the average coordinate can be defined using \re{massless1} for $L(t)=A+Bt$ as
\baa
<x(t)> =\sum_{n,m} a_n^*(0)a_m(0)e^{i\frac{\pi(n-m)}{B}\ln\left( 1+\frac{B}{A}t\right)} V_{nm} \nonumber
\eaa

In Fig. 3(a) the average coordinate is plotted as a function of time for three regimes of wall's motion, for linearly expanding, contracting and harmonically oscillating walls.  Zitterbewegung can be clearly observed for all three regimes.

One of the characteristics of the dynamical confinement is so-called quantum force which was considered, e.g. in the case of  
nonrelativistic
counterpart of our model in \ci{nakamura02}. 
The expectation value of the force, $<F(t)>$ acting on a Dirac particle can be determined as   
\baa
<F(t)> = -<\Psi|\frac{\partial H}{\partial L(t)}|\Psi> = -\frac{\partial }{\partial L(t)} <E(t)>, \nonumber
\eaa
where $<E(t)>$ is average kinetic energy of the particle.
It is easy to see that
\baa
<F(t)>=-\frac{1}{\dot{L}}\frac{\partial }{\partial t}<E(t)> = \sum_{n=0}^\infty F_n(t) \nonumber
\eaa
where 
\baa
F_n(t)=\frac{1}{L(t)^2}(|a_n(\tau(t))|^2 - |a_{-n}(\tau(t))|^2). \nonumber
\eaa
$n\in {\mathbb N}$. Apparently, $F_0(t)=0$.


Fig. 3(b) presents plots of the force $<F(t)>$ as a function of time for three regimes of wall's motion, linearly expanding, contracting and harmonically oscillating ones. The force grows as a function of time for contracting box, while one observe its decay for expanding box. It oscillates for the oscillating wall.

An important case that can be studied for dynamical confinement is its
adiabatic limit, i.e. the case, when position of the wall changes very slowly. An interesting feature of such system is existence of so-called geometric phase.  Berry found \ci{Berry} that for a quantum system evolving adiabatically due to a slowly varying parameter, if the parameter  adiabatic changes along the closed curve, $C$ in the parameter space,  the wave function of the system can acquire the so-called geometrical phase in addition to the dynamical one. This additional phase (called later "Berry phase") is different from zero when the trajectory  of the system in the parameter space is located near a point at which the states of the system are degenerate. 
For the  Dirac particle in a time-dependent box, one can show by direct calculation from Eq.\re{DirMod1} that 
\baa 
\gamma_n(C)=i\oint\left(\int\phi^\dag_n(M(\tau))\partial_M\phi_n(M(\tau))dx\right) dM =0,
\lab{bph1} \nonumber
\eaa 
i.e., there is no Berry phase for Dirac particle with time-dependent mass in one dimension.

\section{Conclusions}
In this work, we studied dynamics of a Dirac particle confined in a one-dimensional box with moving wall. The system is described in terms of the one-dimensional Dirac equation with time-dependent boundary conditions. We developed an approach based on modification of the time-derivative to the case of time-varying domain. Within such approach, Dirac equation can be reduced to that for a Dirac particle with time-dependent mass confined in a time-independent box with unit size. 
%
Its solution is reduced to the system of pairs of ordinary differential equations for arbitrary time-dependence of the wall's position which are solved numerically.
Using the obtained solution, we computed average kinetic energy as a function of time. The analysis of such time-dependence shows that there is no unbounded acceleration of the particle in case of 
periodically
oscillating wall
and below bounded box length in general.
Trembling motion in the system is studied by computing average position for different regimes of the wall's motion. Average quantum force created by a moving wall is also analyzed as a function of time. Absence of the geometric phase in the system's adiabatic evolution is shown. The above model can be used in quantum optics where Dirac particles trapped in optically created boxes appear. Also, applications in condensed matter physics, where different quasiparticles can mimic Dirac fermions could be considered. Dynamics of such quasiparticles  optical traps,  or  in (e.g., graphene) quantum dots with varying boundaries can be modelled using the above results.

\section*{Acknowledgements}
This paper is supported by European Union's Horizon 2020 research and innovation programme under the Marie Sklodowska-Curie grant agreement ID: 873071, project SOMPATY (Spectral Optimization: From Mathematics to Physics and Advanced Technology).

\end{document}